\documentclass[runningheads,a4paper]{llncs}

\usepackage[english]{babel}

\usepackage[%
rm={oldstyle=false,proportional=true},%
sf={oldstyle=false,proportional=true},%
tt={oldstyle=false,proportional=true,variable=true},%
qt=false%
]{cfr-lm}
\usepackage{graphicx}

\usepackage{paralist}

\usepackage{cite}

\usepackage[T1]{fontenc}

\usepackage[math]{blindtext}

\usepackage{csquotes}

\usepackage{microtype}

\usepackage{url}
\makeatletter
\g@addto@macro{\UrlBreaks}{\UrlOrds}
\makeatother

\usepackage{xcolor}

\usepackage[
bookmarks=false,
breaklinks=true,
colorlinks=true,
linkcolor=black,
citecolor=black,
urlcolor=black,
pdfpagelayout=SinglePage,
pdfstartview=Fit
]{hyperref}
\usepackage[all]{hypcap}

\usepackage{pdfcomment}

\usepackage[capitalise,nameinlink]{cleveref}
\crefname{section}{Sect.}{Sect.}
\Crefname{section}{Section}{Sections}

\usepackage{xspace}

\DeclareFontFamily{U}{MnSymbolC}{}
\DeclareSymbolFont{MnSyC}{U}{MnSymbolC}{m}{n}
\DeclareFontShape{U}{MnSymbolC}{m}{n}{
    <-6>  MnSymbolC5
   <6-7>  MnSymbolC6
   <7-8>  MnSymbolC7
   <8-9>  MnSymbolC8
   <9-10> MnSymbolC9
  <10-12> MnSymbolC10
  <12->   MnSymbolC12%
}{}
\DeclareMathSymbol{\powerset}{\mathord}{MnSyC}{180}

\begin{document}

\input glyphtounicode.tex
\pdfgentounicode=1

\title{Highly Scalable and Flexible Model for Effective Aggregation of Context-based Data in Generic IIoT Scenarios
\thanks{This is a preprint of a work published at the 2017 Central European Workshop on Services and their Composition (ZEUS 2017).
Please cite as follows:
S. D. Duque Anton, D. Fraunholz, J. Zemitis, F. Pohl, and H. D. Schotten, : ``Highly Scalable and Flexible Model for Effective Aggregation of Context-based Data in Generic IIoT Scenarios.''
In: Proceedings of the 9th Central European Workshop on Services and their Composition (ZEUS 2017), 2017, pp. 51-58.}
}
\titlerunning{Context-based Data Aggregation Model}

\author{Simon Duque Ant\'{o}n \and Daniel Fraunholz \and Janis Zemitis \and Frederic Pohl \and Hans Dieter Schotten}
\authorrunning{Simon Duque Ant\'{o}n et al.}
\institute{German Research Center for Artificial Intelligence \\ Intelligent Networks Research Group \\ \email{\{firstname.lastname\}@dfki.de}}

%

\maketitle

\begin{abstract}
Interconnectivity of production machines is a key feature of the Industrial Internet of Things (IIoT).
This feature allows for many advantages in producing.
Configuration and maintenance gets easier,
as access to the given production unit is not necessarily coupled to physical presence.
Customized production of goods is easily possible,
reducing production times and increasing throughput.
There are, however, also dangers to the increasing talkativeness of industrial production machines.
The more open a system is,
the more points of entry for an attacker exist.
Furthermore, the amount of data a production site also increases rapidly due to the integrated intelligence and interconnectivity.
To keep track of this data in order to detect attacks and errors in the production site,
it is necessary to smartly aggregate and evaluate the data.
In this paper,
we present a new approach for collecting,
aggregating and analysing data from different sources and on three different levels of abstraction.
Our model is event-centric, 
considering every occurrence of information inside the system as an event.
In the lowest level of abstraction,
singular packets are collected, 
correlated with log-entries and analysed.
On the highest level of abstraction,
networks are pictured as a connectivity graph,
enriched with information about host-based activities.
Furthermore,
we describe our work in progress of evaluating our aggregation model on two different system settings.
In the first scenario, 
we verify the usability of our model in a remote maintenance application.
In the second scenario,
we evaluate our model in the context of network sniffing and correlation with log-files.
First results show that our model is a promising solution to cope with increasing amounts of data and to correlate information from different types of sources.
\end{abstract}


\section{Introduction}\label{sec:intro}
The fourth industrial revolution is a driver that leads to increased interconnectivity of production assets.
Machines and goods are equipped with intelligence as well as communication abilities to negotiate processing steps.
As a result, a lot of communication takes place,
within a production site but also across the boundaries of several facilities.
For several reasons,
such as security and intrusion detection,
pre-emptive maintenance,
quality management and error detection,
it is necessary to monitor this communication data.
The data generated by smart devices also contains relevant information about conditions of entities.
Particularly in Industrial Internet of Things (IIoT), the above mentioned effects take place:
The amount of devices is rising generating more and more of various data,
being increasingly interconnected.
On the other hand, the production of value shifts from physical assets to Intellectual Property (IP),
making the Information Technology (IT) infrastructure of a company a desired target for adversaries~\cite{FancherX}.
Properties like Self-Healing, Self-Protection and Self-Organziation are demanded by industry and require a lot of information~\cite{Neves2016}.
Furthermore, 
the information of controlling entities such as Manufacturing Execution Systems (MES),
Enterprise Resource Planning (ERP) or maintenance ticketing systems can be used to validate the soundness of device behaviour.
Even though a lot of information is already generated and readily available in modern industrial applications,
there are very few means to connect and correlate this data.
In order to gain overview over complex interconnected systems,
the data obtained needs to be made sense of,
not only with respect to the entity it was generated on, 
but in the context of the whole network.
The sensors for generating data are usually already available,
e.g. the Programmable Logic Controller (PLC), router, switch, MES, ERP or something similar,
only the correlation of data has been neglected.
In this paper, 
a holistic approach to collect and connect this data is presented.
In our approach,
every information is derived from time-discrete events.
It therefore contributes to the field of Complex Event Processing (CEP) in the domain of IIoT.
We describe a model to classify the data that can be extracted and a way to connect data from different sources.
This is especially useful to take the context of events into consideration and to semantically assess them.
This paper is organized as follows.
In chapter II,
the state of the art of monitoring and aggregation of data as well as context-aware data analytics is described.
An exhaustive specification of our data aggregation model is presented in chapter III.
In chapter IV, 
the application of our model to two use cases is evaluated.
A conclusion is drawn in chapter VI,
as well as an outlook on the continuation of our work.

\section{Related Work}\label{sec:sota}
Event processing has been brought up in the 1950's in the context of discrete event simulation\cite{2007LUCKE}.
Since then,
a lot of development has taken place.
Nowadays, 
because of the rising amount of data and the increasing complexity of information,
databases and sophisticated queries are employed\cite{2006WU, 2012CUGOLA}.
This field is called "CEP" and applied on many domains, 
often on business and finance intelligence\cite{2008ECKERT, 2009BUCHMANN},
but also on education\cite{2001DODDS} or
automation purposes\cite{2010ROBINS, 2005WANG}.
For network management and IT security,
similar concepts are used under the name of Event Correlation\cite{1993KLIGER, 2013FICCO, 2004JIANG}.
The methods of CEP are used to correlate, aggregate and access information in order to gain intelligence that is only available due to the combination of multiple events\cite{1998LUCKHAM}.
Most CEP systems are based on data bases and tailored query languages,
for example SASE\cite{2007GYLLSTROM},
but many others as well\cite{2012CUGOLA}.
One core idea of CEP is gaining and storing only relevant information from a wide range of sources. 
This is called "aggregation".
One of the current issues of CEP is the horizontal correlation of events\cite{2004JIANG, 2010ROBINS}.
Correlation is used to describe relations of cause and effect. 
This is a non-trivial topic and is heavily domain dependant task\cite{2004JIANG, 2008ECKERT}.
Our model proposes a concept for this issue in the domain of IIoT.

\section{Concept of the Aggregation Model}\label{sec:concept}
In this section,
the motivation for as well as the concept of the aggregation model are described.
As mentioned in section \ref{sec:intro},
modern systems generate a lot of data that is especially useful for anomaly detection.
The biggest issue lies in the distribution of this data: 
Especially in industrial applications,
there are lots of information on local instances of devices,
however only a relatively small part is gathered and presented to a human operator.
The abundance on available information makes it impractical to look at each feature separately.
In order to change this, 
a model is needed to gather data centrally in order to correlate individual events,
generated by different types of data sources and see relationships of different activities.
On the other hand,
it is tedious, computationally expensive and time-consuming to gather all information in a central place and process it.
Today,
most embedded devices are capable of performing complex computations on their own.
The presented model aims at bringing together all relevant information from a heterogeneous network,
as typically found in industrial environments,
while at the same time performing as much computation as possible on the devices that generated it.
This keeps the data that needs to be collected on a global scale small,
while still allowing for a full and expressive picture of the network conditions.
It consists of three vertical and horizontal levels each,
as can be seen in table \ref{tab:am}.
The vertical levels depict the layer of abstraction.
The lowest level in the table corresponds to the highest resolution,
while the highest level corresponds to the most abstract view.
To ascend from a level to another,
data has to be summarized.
The horizontal levels describe the relation between different sources of data.
Each column represents a different kind of data source that can be used to get a specific kind of data.
Three types of data sources are chosen, since three different kinds of data sources in industrial systems were identified.

\begin{table}
\caption{Aggregation Model}
\begin{tabular}{|l|c|c|c|}
\hline
& Cause & Traffic & Effect \\
\hline
Level 1 & e.g. Configuration entries & e.g. Network Packets & e.g. Log-entries \\
\hline
Level 2 & e.g. Application Info & e.g. Network Flow & e.g. Log-files \\
\hline
Level 3 & e.g. Data base & e.g. Flow Graph & e.g. Log file collection \\
\hline
\end{tabular}
\label{tab:am}
\end{table}

The leftmost column describes the cause of an event. 
It is the command or the action on a system that triggers network traffic and an action on a remote system.
In the middle column, 
the network traffic of an event is depicted.
The rightmost column describes the effect of an event.
It is the action on a remote machine,
triggered by some action or the activity a PLC executes due to a change of parameters.
We assume a network-centric approach, 
meaning that only actions that spread within the network are considered relevant.
This is because of the increasing importance of interconnectivity and the inherent effects of it on computing.
Actions on a single host can translated to information of effect and cause, 
without any network traffic.
This case,
however,
is not the goal of our model.
Those kinds of data need to be correlated in order to gain intelligence on the system that is represented by the model.
The general hypothesis is that in an interconnected environment,
no event takes place without some cause, some effect and, since the scope extends to networks,
not only to host based systems, some traffic.
If those three parts of an event can be gathered,
the knowledge about said event consists of all possible points of view,
not only of one systems perspective that could be flawed or tampered.
\newline
\par
The layers are introduced to encode the resolution that is used.
Three layers of resolution are chosen, since they allow for a differentiated view on the system.
The level of abstraction ranges from singular host and network connection to the whole system.
More layers would lead to an increased effort in aggregation without additional benefit.
On the lowest level, Level 1 in table \ref{tab:am}, 
singular network packets are collected and analysed.
They are correlated with singular log-entries on both cause and effect side.
The idea is that a single IP-packet is the smallest unit worth analysing.
On both sides of the communication, 
actions can be determined that either result in or were caused by this traffic,
such as log entries or syscalls.
This information can be aggregated and taken to a higher level of abstraction, Level 2 in table \ref{tab:am}.
The network packets can be aggregated to flows.
Flows only contain source and destination as well as duration,
number of packets some more meta information.
The log-entries on the hosts involved can be aggregated to log-traces,
as can the syscalls.
On the machine labelled as the cause, 
application data can be taken into account,
depending on the application to gather intelligence on the context of the event.
On the highest level of abstraction, Level 3 in table \ref{tab:am},
the whole network is taken into consideration.
The network-flows are aggregated to network topology graphs.
The log- and syscall traces as well as the application information are stored in a compressed form as node-information and correlated with the corresponding edges in the graph.
This allows for a compact,
yet complete overview of the network.
So on the first level, 
a local, possibly extensive analysis of all data takes place, since all information is available.
In order to collect it on a higher level, however, some data reduction needs to take place.
Only relevant information, such as metadata, summarized log/event traces or reduced application settings are stored.
This allows for a complete, yet compact representation of the whole system on level 3.

\section{Use Cases and Evaluation}\label{sec:uce}
The proposed model can easily be applied to different use cases, since it is not domain specific but abstract. 
Even though in this work, the main focus is on IIoT,
the proposed model can be applied for almost any domain that consists of more than one system.
Furthermore, it is highly scalable, as on each level, the number of sensors can be extended without needing to change the overall structure.
Still, the amount of data that is gathered on each level will be manageable due to the principle of aggregation.
In this chapter,
two use cases of our model are depicted.
The first use case is a remote maintenance scenario.
In the context of the second use case,
modbus-based network traffic is correlated with logs gathered from PLCs.
Although the model can be used for a multitude of applications,
our main focus is IT-security by means of anomaly detection in network behaviour.
In both use cases,
the data is collected and analysed in order to detect malicious behaviour.

\subsection{Remote Maintenance}
\begin{figure}
\centering
\includegraphics[scale=0.6]{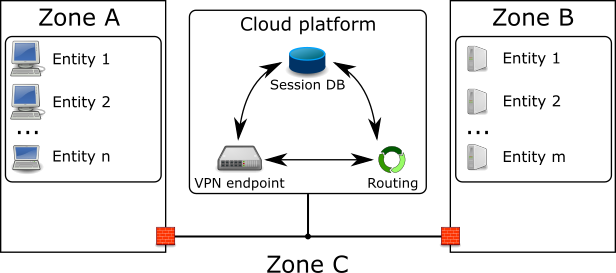}
\caption{Architectural Overview Remote Maintenance}
\label{fig:aorm}
\end{figure}

In increasingly interconnected production networks,
secure remote maintenance gains importance.
We apply our model to an architecture for secure session-based remote administration and maintenance that suits the requirements of segmented and firewalled production network environments.
The architecture is depicted in fig. \ref{fig:aorm}\footnote{The icons have been taken from\cite{OSSARCH}}.
It is divided by means of the zones A, B with limited,
i.e. firewalled,
access to zone C,
thus without the ability to setup zone-spanning peer-to-peer connections.
The cloud platform takes the role of a relay agent,
forwarding legitimate traffic between entities from zone A to zone B and vice versa.

The cloud platform consists of three core components,
each of them contributing information to the aggregation model.
The session database holds information about scheduled,
ongoing as well as terminated remote administration sessions.
It is queried by the VPN endpoint in order to additionally authenticate connecting entities,
as well as the routing engine to setup forwarding tables according to session properties.
Any traffic related to remote administration passes the VPN endpoint and can easily be extracted for aggregation.
Additional session state information can be obtained by the routing engine.
We further note that communication between the core components may also be subject to aggregation. \\ \par
The process of a remote maintenance action starts after the machine owner in zone A notices some malfunction in her machines.
She issues a maintenance ticket to the cloud platform.
A maintenance technician accepts the ticket, is authorized to access the machine in the context of the maintenance process and performs the maintenance action.
After the technician is done, she disconnects and the machine revives operation.
In our implementation, iptables are employed on the cloud platform in order to re-route the clients via a OpenVPN connection.
The ticket handling is done with OTRS.
In this use case, the data collection is machine owner-centric, 
this means we assume that the machine owner only has access to her own network in Zone A.
Each ticket, that has been issued, is collected as a level 1 cause. 
They can be used to provide insight about the condition of a machine.
Each setting of the machine after the maintenance action took place is collected as a level 1 effect.
The traffic between the machine and the maintenance technician is the level 1 traffic.
This information can be gathered and processed locally. 
It is then aggregated and collected on level 2,
where the network traffic is abstracted to a network flow,
the tickets of a singular machine are summarized to a ticket history as the machine settings after the maintenance are concentrated to setting histories. 
On the level 3, this information is combined with information from different machines of the machine owner. \\ \par
On level 1, the individual network packets are correlated with log entries of the machine.
This can help in detecting inconsistencies in the programming, 
for example if a ticket is issued even though no error or warning could be detected in the machine logs.
Another possibility lies in an maintenance technician to misconfigure a machine.
The comparison of current network traffic and machine settings to the previous settings of the machine and the corresponding traffic on level 2 can detect such errors.
On level 3, all machines of the same owner are considered. 
This way, machines that have a high rate of errors can be identified,
as this can be a sign of adversarial activities.

\subsection{Log and traffic correlation}
\begin{figure}
\centering
\includegraphics[scale=0.6]{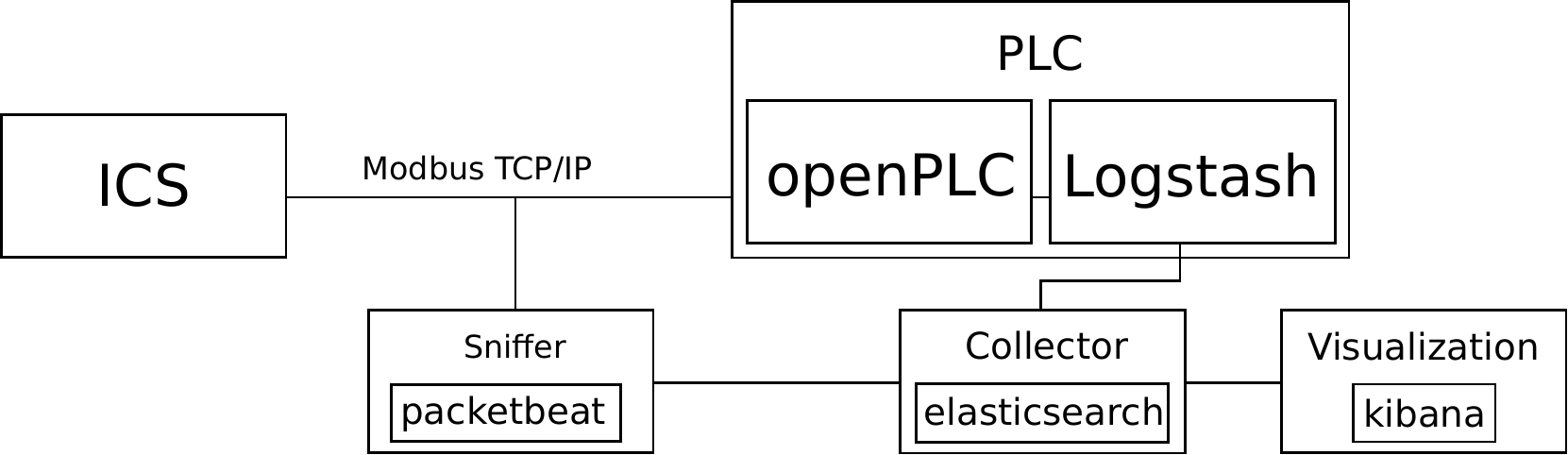}
\caption{Architectural Overview PLC Aggregation}
\label{fig:aopa}
\end{figure}

PLCs are computation units in industrial networks that control production processes.
They can be configured via network access.
In order to gain intelligence,
we propose to collect network traffic data as well as host-based log information on the PLC.
The schematic architecture that helps us gather the information necessary is pictured in fig. \ref{fig:aopa}.

The Industrial Control System (ICS) communicates with the PLC we set up using openPLC \cite{OPENPLC}.
The network traffic is gathered with packetbeat,
the log inforamtion can be extracted with logstash.
Elasticsearch serves as a collector,
the findings can be visualized with kibana.
These tools are provided by the company elastic \cite{ELASTIC}. \\ \par
Every change on the PLC is issued by the ICS and transmitted via a fieldbus protocol.
On level 1, the settings of the application on the ICS, the individual network packets and the log entries and register settings of the PLC are collected.
Inconsistencies, such as parameterization from an ICS that has no knowledge about this process,
are a hint for malicious activities, e.g. spoofing~\cite{Tanase2003}.
The abstraction to level 2 allows for a more abstract view of all past settings in order to capture unlikely or unprecedented behaviour that is a sign of erroneous behaviour.
Level 3 allows for the comparison of PLCs and ICSs on the system level,
identifying ICSs that communicate with suspiciously few PLCs or PLCs that are set to singular, unlikely settings.

\section{Conclusion and Outlook}
In this paper,
we introduced a novel data aggregation and correlation model.
We showed that it can be employed in multiple fields by presenting two different use cases.
Furthermore,
we spotlighted features of our model that meet the demands in IIoT.
Those features can be used to help managing the rising amount of data of interconnected and intelligent systems.
Several different tasks,
such as IT-security or maintenance,
motivate the necessity for sophisticated data management.
Our model is able to support this increasingly complex task.
\newline
\par
The contribution of this work is a model that can be used to address the problem of the increasing amount of heterogeneous data,
especially in the industrial environment.
At the same time, industrial applications are getting more attractive for attackers,
creating the need for efficient detection of malicious behaviour.
Correlating and evaluating data is a crucial task in this arising s,
one that is aided by the presented model.
\newline
\par
As a next step,
we will develop connection operators to make correlation of events easier.
We will enrich our list of data sources to get a more holistic overview of networks.
Other than that,
we plan to incorporate algorithms of machine learning into our framework to help the system find correlations between events and detect anomalies of any kind.

\subsubsection*{Acknowledgments}
This work has been supported by the Federal Ministry of Education and Research of the Federal Republic of Germany (Foerderkennzeichen KIS4ITS0001, IUNO). The authors alone are responsible for the content of the paper.


\bibliographystyle{splncs03}
\bibliography{paper}

\begin{thebibliography}{10}
\providecommand{\url}[1]{\texttt{#1}}
\providecommand{\urlprefix}{URL }

\bibitem{ELASTIC}
{elastic}. \url{https://www.elastic.co/de/}

\bibitem{OPENPLC}
{OpenPLC}. \url{http://www.openplcproject.com/}

\bibitem{OSSARCH}
{OpenSecurityArchitecture}.
  \url{http://www.opensecurityarchitecture.org/cms/library/icon-library}

\bibitem{2009BUCHMANN}
Buchmann, A., Koldehofe, B.: {Complex Event Processing}. In: it - Information
  Technology. vol.~5. Oldenbourg Wissenschaftsverlag (2009)

\bibitem{2012CUGOLA}
Cugola, G., Margara, A.: {Processing Flows of Information: From Data Stream to
  Complex Event Processing}. In: ACM Computer Surveys (2012)

\bibitem{2001DODDS}
Dodds, P., et~al. (eds.): Sharable Content Object Reference Model (SCORM(TM)).
  Advanced Distributed Learning Initiative (2001)

\bibitem{2008ECKERT}
Eckert, M., Bry, F.: {Aktuelles Schlagwort: Complex Event Processing (CEP)}.
  Springer Verlag (2008)

\bibitem{FancherX}
Fancher, D.: {5 insights on cyberattacks and intellectual property}.
  \url{https://www2.deloitte.com/content/dam/Deloitte/us/Documents/finance/us-fas-five-insights-on-cyber-attacks-and-intellectual-property.pdf}

\bibitem{2013FICCO}
Ficco, M.: {Security Event Correlation Approach for Cloud Computing}. In:
  International Journal of High Performance Computing and Networking (2013)

\bibitem{2007GYLLSTROM}
Gyllstrom, D., Diao, Y., Wu, E., Stahlberg, P., Chae, H.J., Anderson, G.:
  {SASE: Complex Event Processing over Streams} (2007)

\bibitem{2004JIANG}
Jiang, G., Cybenko, G.: {Temporal and Spatial Distributed Event Correlation for
  Network Security}. In: Proceedings of the American Control Conference (2004)

\bibitem{1993KLIGER}
Kliger, S., Yemini, S., Yemini, Y., Ohsie, D., Stolfo, S.: {A Coding Approach
  to Event Correlation} (1993)

\bibitem{2007LUCKE}
Luckham, D.: {A Short History of Complex Event Processing} (2007)

\bibitem{1998LUCKHAM}
Luckham, D.C., Frasca, B.: {Complex Event Processing in Distributed Systems}
  (1998)

\bibitem{2010ROBINS}
Robins, D.B.: {Complex Event Processing} (2010)

\bibitem{Neves2016}
Santos, J.P., Alheiro, R., Andrade, L., Ángel Leonardo Valdivieso~Caraguay,
  López, L.I.B., Monge, M.A.S., Villalba, L.J.G., Jiang, W., Schotten, H.D.,
  Calero, J.M.A., Wang, Q., Barros, M.J.: {SELFNET Framework Self-Healing
  Capabilities for 5G Mobile Networks}  (2016)

\bibitem{Tanase2003}
Tanase, M.: {IP Spoofing: An Introduction}.
  \url{https://www.symantec.com/connect/articles/ip-spoofing-introduction}
  (2003)

\bibitem{2005WANG}
Wang, F., Liua, S., Liu, P., Bai, Y.: {Bridging Physical and Virtual Worlds:
  Complex Event Processing for RFID Data Streams} (2005)

\bibitem{2006WU}
Wu, E., Diao, Y., Rizvi, S.: {High-Performance Complex Event Processing over
  Streams} (2006)

\end{thebibliography}

All links were last followed on January 3, 2016.

\end{document}